# Spatially explicit feature importance for building height estimation using research-access high-resolution SAR and optical sensors

1st Guilherme Iablonovski
*Université Gustave Eiffel, Géodata Paris, IGN, LASTIG, F-77454 Marne-la-Vallée, France*
*Programa de Pós-Graduação em Sensoriamento Remoto, Universidade Federal do Rio Grande do Sul, Porto Alegre, Brazil*
orcid.org/0000-0003-0688-3062

2nd Pierre-Louis Frison
*Université Gustave Eiffel, Géodata Paris, IGN, LASTIG, F-77454*
Marne-la-Vallée, France
orcid.org/0000-0002-0956-0085

3rd Tatiana Silva da Silva
*Programa de Pós-Graduação em Sensoriamento Remoto, Universidade Federal do Rio Grande do Sul*
Porto Alegre, Brazil
orcid.org/0000-0002-7234-0042

***Abstract*—Accurate building height information at the individual footprint scale is essential for material stock accounting and post-disaster damage assessments yet remains difficult to obtain at city scale in the Global South where airborne LiDAR coverage is rare and commercial very high-resolution imagery is cost-prohibitive or unavailable. While recent works have demonstrated building height estimation using freely available Sentinel imagery, the resolution ceiling of resulting products is still coarse for material stock analysis. This study incorporates products derived from data freely accessible under scientific research licenses, TerraSAR-X StripMap and PlanetScope, alongside Sentinel-1 to predict building heights in a large city in Brazil. To account for the spatial autocorrelation in the training set, features from all sources are integrated in a geographically weighted random forest model, returning an RMSE of 5.34 m and $R^2$ of 0.756 against a LiDAR reference dataset. Local feature importance showed predictor dominance to vary consistently across intra-urban contexts, with footprint geometry dominating for low-rise buildings, shadow-derived height for taller and more isolated structures, and spectral reflectance for the tallest buildings in the set. Sentinel-1 backscatter and InSAR occupy complementary spatial niches, with no single sensor uniformly preferable across the set. Results provide optioneering guidance and insight over satellite-derived products predictive relevance in distinct contexts, which global machine learning or neural network models cannot offer.**

## I. Introduction

Three-dimensional building information is essential for a wide range of applications such as assessing urban damage following disasters. The extraction of building height data has been approached using many methods in remote sensing over time, including airborne laser scanning (ALS), optical imagery shadow analysis, interferometric SAR, and stereophotogrammetry. Yet, many of the available methods rely on data that imply high costs, such as airborne LiDAR and very high resolution (VHR) satellite imagery, limiting their temporal and spatial scale, coverage, and consistency [1]. This scenario is particularly relevant for the Global South and regions with ongoing conflicts, where ALS exercises are rarer, and VHR sensor archives are often incomplete as they depend on commercial activations.

In this context, pipelines that integrate SAR and optical open-access satellite-derived information with systematic acquisitions have emerged as a path to filling the three-dimensional building data gap. [1] combined Sentinel-1 and Sentinel-2 time series to estimate building height at 10 m spatial resolution across Germany through a support vector machine model, achieving an RMSE of 6.07 m. [2] achieved an RMSE of 1.89 m by advancing this method with a deep learning architecture using a similar dataset. Although useful, results operating at 10 m pixel resolution do not allow for resolve of values for individual building footprints, which is a significant limitation for applications such as building material stock requiring archetype-based material intensities.

In this paper, we propose an approach including not only data free in an absolute commercial sense, but also imagery accessible to academics at scale for extensive geographic coverage. We selected TerraSAR-X (TSX) data, available at no cost through the DLR science program, and PlanetScope optical imagery, available through Planet's Education and Research Program, which provide medium-high spatial resolution and allow for additional data-derived products. We predict building heights for Porto Alegre, a large Brazilian city, using features across seven predictor groups that are compared to reference LiDAR data. Beyond height prediction, we seek to differentiate each method's performance across construction typologies and intra-urban contexts, which is directly relevant to optioneering pipelines operating under constrained budgets and data availability. To this end, features are integrated in a geographically weighted random forest (GWRF) regression. Urban morphology variables are then examined to characterize the urban contexts and building types each source predicts best, providing an interpretable physical account of the variation in predictor importance.

## II. Methods

### A. Reference Dataset creation

*a) Building Height Reference Dataset:* Building footprints and reference heights for Porto Alegre are derived from airborne LiDAR data acquired by the Porto Alegre city hall [3], with a minimum point density of 2.2 points per m² and a mean error of 7.5 cm. We calculate building reference height as the difference between the derived digital surface model (DSM) and digital terrain model (DTM) at each building footprint.

*b) PlanetScope Spectral and Texture Features:* PlanetScope [4] 3m, 4-band surface reflectance composites were used to derive a set of spectral and texture features for each building. Spectral indices included the Normalized Difference Vegetation Index (NDVI), a proxy Normalized Difference Built-up Index (NDBI), and brightness. Gray-level co-occurrence matrix (GLCM) texture features were computed from the near-infrared band, and yielded contrast, entropy, energy, correlation, dissimilarity and homogeneity.

*c) Optical Shadow-derived Heights:* The same optical composites were used to extract shadow geometry as a proxy for building height. Shadow pixels were identified using a shadow index [5], in which values over 0.85 correspond to pixels that are dark across all bands. Shadow length was converted to an estimated building height using the scene's solar incidence angle and azimuth as proposed by [6]. Results were capped to a range (3–110m) to remove implausible estimates arising from shadow occlusion or adjacent-building contamination. A binary availability flag was retained alongside the continuous height estimate.

*d) InSAR-derived Heights:* A TSX StripMap [7] interferometric pair (ambiguity height of 177 m) was processed in ESA SNAP [8] to derive a DSM. To ensure consistency with the criterion motivating the sensor choices in this study, StripMap mode, which has more systematic acquisitions and wider swath width, was selected over other higher-resolution modes. Processing consisted of coregistration, interferogram formation, Goldstein phase filtering ($\alpha = 0.2$), phase unwrapping, and terrain correction. The resulting InSAR DSM was subtracted from the resampled FABDEM DTM [9] to produce an nDSM.

*e) SAR Backscatter and Texture:* Sentinel-1[10] Ground Range Detected (GRD) imagery complements TSX SAR data as it offers a distinct scattering mechanism scale. At Sentinel-1's 20 m spatial scale, double-bounce returns integrate wall-ground interactions from multiple buildings into a single pixel, producing a neighborhood-level volumetric signal [11][12]. All images available for the reference year of 2015 covering the area of interest were processed to create the VVH indicator [12] which is a combination of co- and cross-polarized backscatter sensitive to the double-bounce scattering mechanism, shown to be able to estimate mean building height at 500 m [12]. From the resulting VVH backscatter time series, eleven statistical temporal metrics (STMs) were computed per polarization: mean, standard deviation, the 10th/25th/50th/75th/90th percentiles, range, interquartile range, skewness, and kurtosis. Analogous to the processing done to the optical imagery, seven GLCM texture features were computed on the temporal mean backscatter.

*f) Footprint and Geometric Attributes:* A set of attributes was computed from the building footprint polygons, namely area and perimeter. Isolation metrics were also computed: distance to the nearest neighboring building, mean distance to the five nearest neighboring buildings, and counts of neighboring buildings within fixed radii of 25m, 50m, and 100m. The inclusion of these features is motivated both by their role in characterizing remote sensing signal quality (e.g., shadow and SAR backscatter contamination from adjacent structures) and by recent evidence that footprint morphology and spatial context alone carry substantial predictive signal for building height (e.g. [13]).

For each set of features, zonal statistics per building footprint were computed using an all-touched extraction rule to ensure adequate pixel coverage for small footprints. The final dataset is a vector database containing the features from the different variables, which are illustrated in Figure 1.

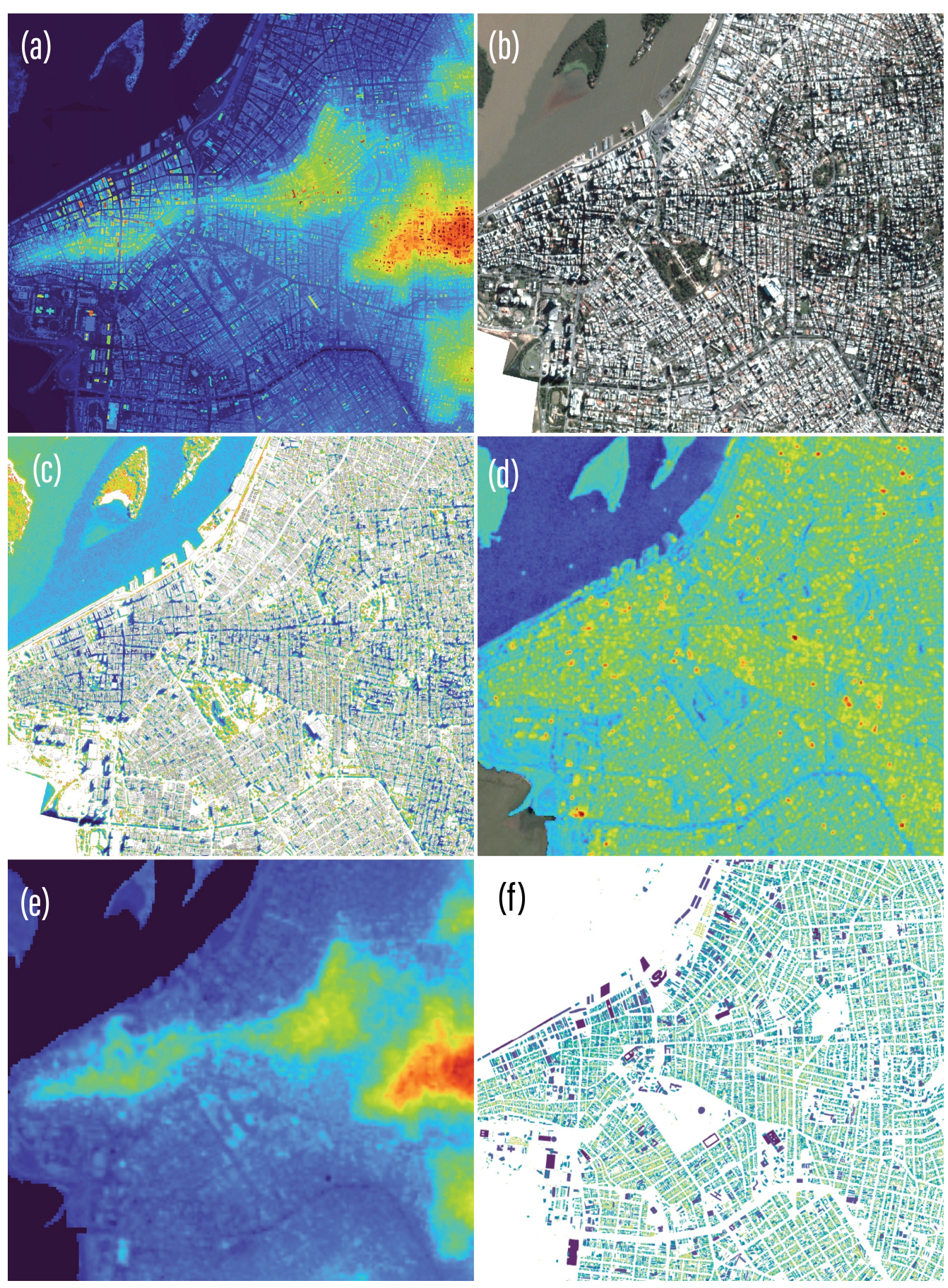

Figure 1. Used sources in their native scales before introduction into footprints through zonal statistics. (a) LiDAR-derived DSM, (b) PlanetScope natural color composite, (c) PlanetScope-derived shadow index, (d) Sentinel-1 VVH-derived backscatter time series median, (e) TerraSAR-X inSAR-derived

## B. Statistical Analysis

*a) Preliminary analysis:* The final assembled dataset contains 1 dependent variable (LiDAR-derived building height), and 66 variables across seven explanatory groups: optical spectral, optical texture, shadow-derived heights, inSAR-derived heights, backscatter intensity, and backscatter texture. To determine the appropriate statistical model to analyse the predictive potential of each group, exploratory analysis was performed. Spatial autocorrelation in building height was assessed via Moran's I on the full building population ($n = 485{,}334$), using a k-nearest-neighbor weights matrix ($k = 8$). The result ($I = 0.374$, $z = 554.18$, pseudo $p = 0.001$ over 999 permutations) confirmed that building heights cluster significantly, motivating a locally-sensitive approach.

Geographically weighted regression (GWR) was considered but discarded as the dataset violates the normality and multicollinearity assumptions. Random forest (RF) regression, alternatively, is unaffected by these concerns but treats all observations as spatially exchangeable. To ensure robustness to the data quality and distribution characteristics and sensitiveness to spatial structure, a geographically weighted random forest (GWRF) was adopted.

*b) Geographically weighted Random Forest:* GWRF [14] extends the random forest algorithm with the spatial weighting logic of GWR: for each building, a local RF is fit using nearby samples weighted by proximity, and final predictions combine local and global models. GWRF bandwidth and the relative weight given to local versus global predictions were informed by an incremental spatial autocorrelation analysis.

Because building heights larger than 12m are rare (4% of stock) yet disproportionately harder to predict than short buildings (whose height is well indicated by footprint alone), we built a spatially stratified sample (33,212 buildings) guaranteeing inclusion of every building 12m or taller, filling remaining positions with a representative draw of buildings across a regular grid to preserve geographic coverage (Table 1).

TABLE I. BUILDING HEIGHT DISTRIBUTION IN THE FULL DATASET POPULATION AND THE GWRF SAMPLE

| Height stratum | Full population (n) | Full population (%) | GWRF sample (n) | GWRF sample (%) |
|---|---|---|---|---|
| <6m | 339,382 | 79.8% | 7,862 | 23.7% |
| 6–12m | 69,012 | 16.2% | 8,334 | 25.1% |
| 12–20m | 10,493 | 2.5% | 10,493 | 31.6% |
| 20–35m | 4,723 | 1.1% | 4,723 | 14.2% |
| >35m | 1,800 | 0.4% | 1,800 | 5.4% |
| **Total** | **425,410** | **100%** | **33,212** | **100%** |

An alternative design would compare residuals from separate per-group models to test which predicted each building best, but this only reflects each group's standalone power, not its marginal contribution. A group could fit a building well for unrelated reasons (e.g., a larger number of features), and this approach can't reveal whether one group's apparent power is redundant with another's once both are available. As the question is which predictor most explains height at a given location given everything else the model has access to, feature importance was assessed within a single joint model rather than via independent per-group models.

Feature importance in RF reflects how much a predictor contributes to the predictions based on how often it is used to split the data across decision trees. It does not, however, indicate whether its relationship with height is positive or negative. In GWRF, importance is computed separately for each building's local model, allowing the predictor that matters most to vary from one location to another. Local feature importance was aggregated by group mean to avoid bias toward groups with more constituent features.

### C. *Urban morphology features*

To characterize which types of buildings and urban contexts are best described by each predictor group, a set of seven morphology and contextual variables was computed for each building: altitude and slope (derived from the LiDAR-based digital terrain model), footprint orientation and elongation (derived from each footprint's minimum rotated bounding box), footprint vertex count, neighborhood built density, and height relative to the mean of the five nearest neighboring buildings. These were compared across the seven dominant local-importance groups using one-way ANOVA, with Tukey's HSD post-hoc tests used to identify which group pairs differed significantly.

## III. RESULTS

GWRF was evaluated via 10-fold cross-validation on the full 66-feature set (n = 33,212). The combined model achieved RMSE = 5.34 m and $R^2$ = 0.756, outperforming both the global-only component (RMSE = 5.76 m, $R^2$ = 0.716) and the local-only component (RMSE = 5.53 m, $R^2$ = 0.739), confirming that spatial adaptivity improves prediction across the urban fabric. The spatial distribution of feature group importance is shown in Figure 2 for selected groups displaying clear importance clusters.

Across the full sample, attributes from the geometry group were the dominant predictor for the largest share of buildings (32.4%), followed by shadow-derived height (24.5%), PlanetScope spectral reflectance (16.0%), PlanetScope texture (11.9%), InSAR-derived height (9.0%), Sentinel-1 texture (3.5%), and Sentinel-1 backscatter (2.7%). This ranking differs substantially from the picture given by a global, non-spatial and non-height-stratified random forest, in which geometric attributes alone accounted for the dominant predictor in 94.5% of buildings.

The distribution of dominant predictor groups also varied systematically with building height, as shown in Figure 3, where for each building, the group with the highest mean importance is identified as its dominant predictor. Geometry's share declined from 42.9% of buildings under 6 m to 17.7% of buildings above 35 m, consistent with [13] finding that footprint morphology leaves more residual error for taller constructions. Shadow-derived height's share rose sharply between the shortest stratum (4.5%) and all taller strata (36%), consistent with a height threshold below which cast shadows are too short to provide a reliable signal. Shadow-based extraction succeeds for large, isolated, irregularly-shaped structures with sufficient surrounding clear ground, with shadow-dominant buildings being significantly taller relative to their neighbors.

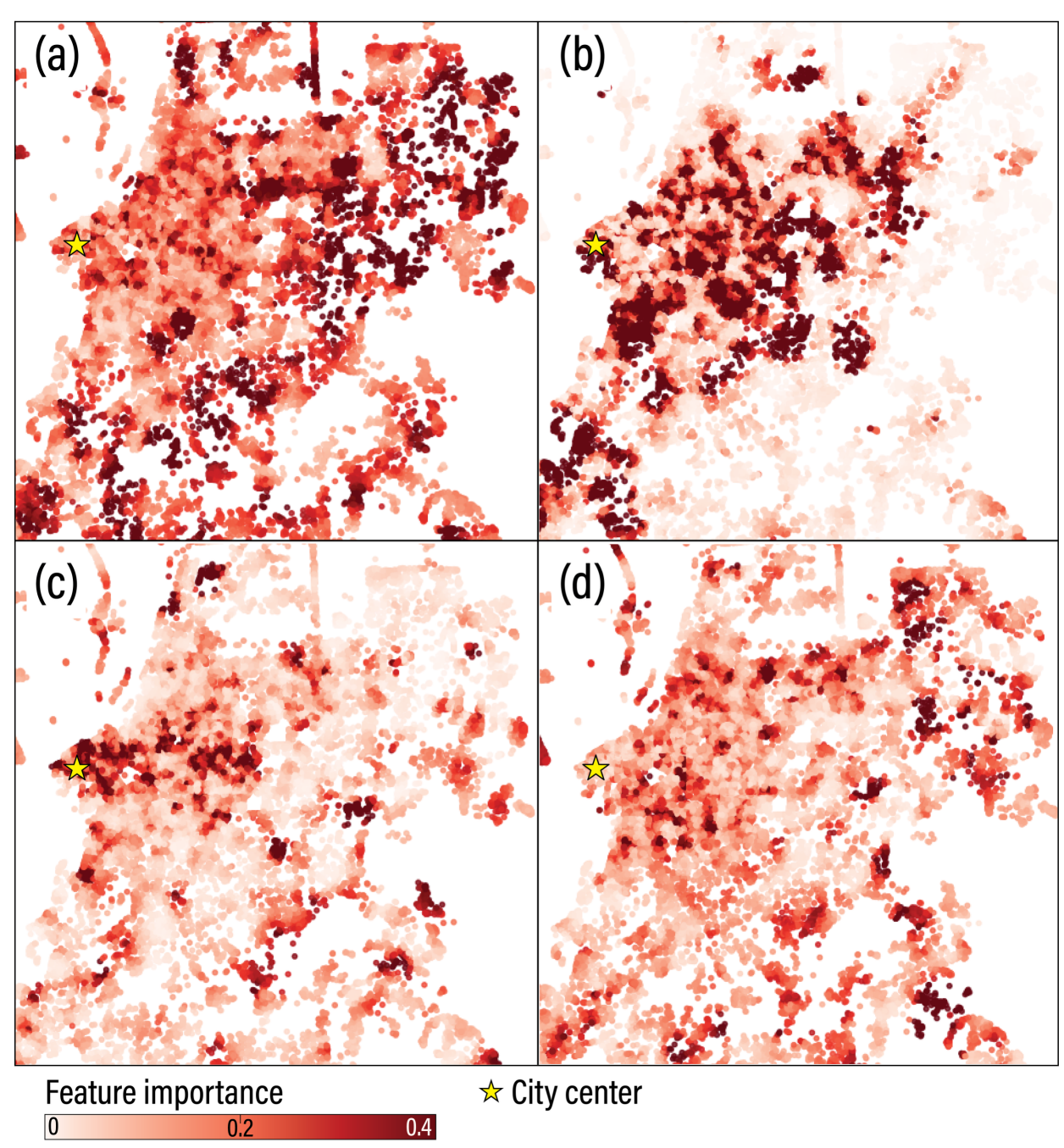

Figure 2. Spatial distribution of feature group importance in the GWRF model. (a) Geometry, (b) Shadow-derived height, (c) PlanetScope spectral, (d) InSAR-derived heights. Source: authors.

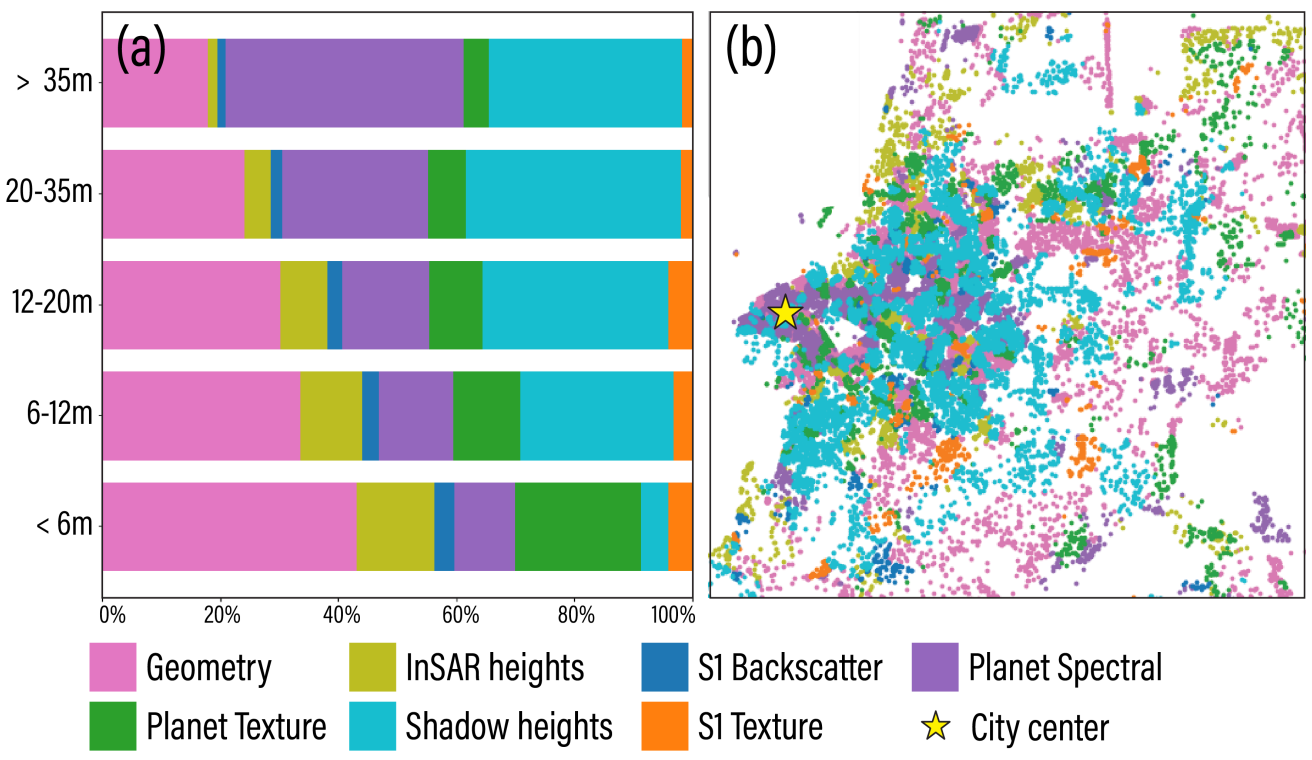

Figure 3. Distribution of dominant feature predictor group by (a) building height stratum and (b) across city fabric. Source: authors.

InSAR-derived height dominated specifically for buildings at low altitude and slope. The InSAR-derived nDSM is computed by differencing the interferometric DSM against FABDEM, which is more prone to mismatching ground surface in complex or rapidly varying terrain, regardless of the DSM's own accuracy. This reflects a methodological artifact rather than a building-level physical mechanism.

PlanetScope spectral reflectance showed the largest overall change across height strata, rising from 10% to 40% and becoming the largest group among the tallest buildings. Buildings where spectral reflectance was dominant were located in significantly denser environments than the other five groups. Very tall buildings in Porto Alegre, as in most cities, are disproportionately commercial or institutional, and differ systematically from the low-rise residential buildings that dominate the city in roofing material, impervious surface coverage, and surrounding vegetation, all of which the spectral indices used here are sensitive to.

Neither SAR nor shadow-derived features maintain their discriminative power as height increases, consistent with the saturation effects reported by [1]. The observed compensation of diminishing height sensitivity of inSAR and shadow geometry with correlated spectral signatures indicates that future work incorporating additional sensing modalities or architectures would offer more significant gains if focus is given to taller (>35 m) buildings.

One-way ANOVA confirmed that all twelve morphology variables tested differed significantly across the seven dominant-group categories (all $p < 0.001$), confirming that the spatial heterogeneity hypothesis holds across multiple independent morphological dimensions.

## Conclusion

This study demonstrates that building height estimation at the individual footprint scale is achievable using only remote sensing data accessible at no cost to researchers. The spatial patterns of predictor dominance provide actionable optioneering guidance that a global, pixel-level model cannot offer, and confirms that sensor contributions vary systematically across typologies and urban contexts.

These results extend the open-data rationale established by [1] and [2] by demonstrating that sensors available under scientific research licenses contribute spatial resolution sufficient for native footprint-level prediction. The study also demonstrates that footprint-level height estimation is feasible in a Global South context where LiDAR coverage is rare and VHR commercial archives are incomplete.

It is worth acknowledging that the instance-level RMSE of 1.60 m reported by [2] substantially outperforms the 5.34 m reported here, which suggests that a deep learning architecture trained on large multi-country datasets still offers superior accuracy. The use of GWRF in this study is therefore not primarily motivated by accuracy but by interpretability. Unlike deep learning models, GWRF can produce spatially explicit feature-importances that are directly informative for future optioneering in resource-constrained contexts.

Future work should examine whether the spatial patterns identified here generalize to other cities, and if the addition of products improves performance in the tall-building range.

## Acknowledgment

TerraSAR-X data were provided by the European Space Agency under the Third Party Missions programme (project PP0109005). PlanetScope imagery was provided by Planet Labs PBC through the Education and Research Program. The authors thank ESA Earth Observation User Services and Planet Labs PBC for data access.